\documentstyle[amssymb,prl,aps,epsfig]{revtex}

\begin{document}

\title{Relationship between the isotope effects on transition temperature,
specific heat and penetration depths}
\author{T. Schneider \\
Physik-Institut der Universit\"{a}t Z\"{u}rich, Winterthurerstrasse 190,\\
CH-8057 Z\"{u}rich, Switzerland}
\date{}
\maketitle

\begin{abstract}
We show that in anisotropic superconductors, falling at finite
temperature into the 3D-XY- and at zero temperature into the
2D-XY-QSI universality class, the isotope effects on transition
temperature, specific heat and magnetic penetration depths are not
independent, but related by universal relations. The corresponding
experimental data for cuprate superconductors reveals suggestive
consistency with these relations. They imply a dominant role of
fluctuations so that pair formation and pair condensation do not
occur simultaneously. For this reason and due to the occurrence of
the 2D-XY-QSI transition the isotope effects do not provide
information on the underlying pairing mechanism but must be
attributed to a shift of the phase diagram upon isotope
substitution. The 2D-QSI transition is driven by the exchange of
pairs, which favors superconductivity and the combined effects of
disorder and Coulomb repulsion of the pairs, which favor
localization. Since isotope substitution will hardly affect
disorder and Coulomb repulsion, the shift of the phase diagram
must be attributed to the electron-phonon interaction, which
renormalizes the mass of the fermions and with that the mass and
the exchange of the pairs.
\end{abstract}

\bigskip

\bigskip
\section{Introduction}

Isotope effect measurements were of crucial importance in
establishing the phonon-induced BCS model of
superconductivity\cite{bardeen}. In the weak-coupling BCS model
the isotope effect derives from the superconducting transition
temperature $T_{c}$ being proportional to the Debye temperature.
Since the Debye temperature is proportional to $M^{1/2}$ , where
$M$ is the atomic mass, the isotope-effect coefficient for the
transition temperature is $\beta _{T_{c}}=$ $-d\ln
T_{c}/dlnM=0.5$. The later inclusion of Coulomb repulsion in the
pairing interaction resulted in predicted values of less than
$0.5$ and enabled a reasonably good fit to the
data\cite{gladstone}. The situation is more complex in cuprate
superconductors, where $\beta _{T_{c}}$ is small near optimal
doping and maximum $T_{c}$ and large close to the quantum
superconductor to insulator (QSI) transition where $T_{c}$
vanishes\cite{franckrev,tshksing,tshkprl}. Moreover, recently it
has been shown that the effect of isotope substitution is not
restricted to $T_{c}$, but affects the in-plane penetration depth
as well\cite {tshkprl,zhaon,zhao,hoferprl,khasanov}.

\begin{figure}[tbp]
\centering
\includegraphics[totalheight=6cm]{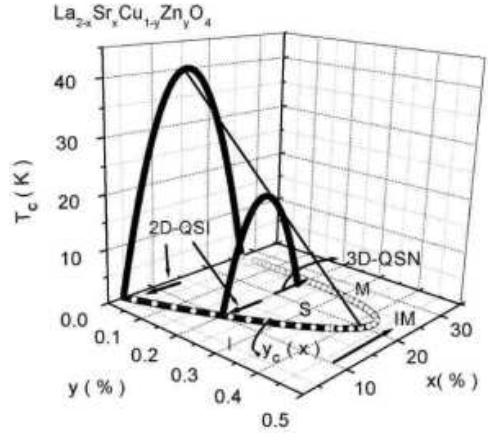}
\caption{$T_{c}\left( x,y\right) $ phase diagram of
La$_{2-x}$Sr$_{x}$Cu$_{1-y} $Zn$_{y}$O$_{4}$ derived from the
resistivity data of Momono \emph{et al}. \protect\cite{momono} and
Fukuzumi \emph{et al}.\protect\cite{fukuzumi}. $y_{c}\left(
x\right) $ is a line of quantum phase transitions. The arrows mark
two dimensional quantum superconductor to insulator (2D-QSI,
$x\lesssim 0.19$), three dimensional quantum superconductor to
normal state transition (3D-QSN,$\ x\gtrsim 0.19$) and insulator
to normal state (IM) crossover, respectively.} \label{fig1}
\end{figure}

The main purpose of this paper is to demonstrate that this
unconventional behavior of the isotope coefficients of $\ T_{c}$,
in-plane penetration depth, etc. is caused by the dominant role of
disorder, thermal and quantum fluctuations and not related to the
pairing mechanism. This differs fundamentally from conventional
superconductors, where fluctuations do not play an essential role
so that the formation of pairs and their condensation occur
essentially simultaneously and where the isotope effect allowed to
identify the electron-phonon interaction as the dominant pairing
interaction. Our starting point is the phase transition surface
$T_{c}\left( x,y,M\right) $ where $x$ is related to the hole
concentration , $y$ denotes the substituent concentration, e.g.
for Cu, or the strength of disorder and $M$ is the isotope mass.
Cuprates where these parameters can be varied include La$_{2-x}
$Sr$_{x}$Cu$_{1-y}$A$_{y}$O$_{4}$ (A= Ni, Zn,...) and
Y$_{1-x}\Pr_{x}$Ba$_{2} $Cu$_{3}$O$_{7-\delta }$, using for Cu
pure $^{63}$Cu or $^{65}$Cu, or for O pure $^{16}$O or $^{18}$O.
In Fig.\ref{fig1} we displayed the phase transition surface
$T_{c}\left( x,y\right) $ for
La$_{2-x}$Sr$_{x}$Cu$_{1-y}$Zn$_{y}$O$_{4}$. $T_{c}$ vanishes for
$y=0$ in the so called underdoped limit $x=x_{u}$, increases for
$x>x_{u}$ and reaches its maximum value at optimum doping
$x=x_{m}$. For any doping level where superconductivity occurs
$x_{u}\leq x\leq x_{m}$ the transition temperature is reduced upon
$Zn $ substitution and vanishes along the line $y_{c}\left(
x\right) $ of quantum phase transitions. The temperature
dependence of the resistivity reveals that along this line two
dimensional quantum superconductor to insulator (2D-QSI)
transitions occur for $x\lessapprox 0.19$, while for $x\gtrapprox
0.19$ three dimensional quantum superconductor to normal state
(3D-QSN) transitions take place\cite{momono,fukuzumi}. Here we
concentrate on the underdoped regime, $x_{u}\leq x\leq x_{m}$.
Since the anisotropy, defined as the ratio $\gamma =\xi _{ab}/\xi
_{c}$ of the correlation lengths parallel $\left( \xi _{ab}\right)
$ and perpendicular $\left( \xi _{c}\right) $ to CuO$_{2}$ layers
(ab-planes) and evaluated at $T_{c}$, tends to diverge by
approaching the endline $y_{c}\left( x\right) $, a three to two
dimensional (3D-2D) crossover occurs\cite{book,tsparks}. Moreover,
the order parameter is a complex scalar, corresponding to a two
component vector (XY). Accordingly, for $x_{u}\leq x\lessapprox
0.19$, $y_{c}\left( x\right) $ is a line of 2D-XY-QSI transitions.
On the other hand, there is mounting evidence that at finite
temperature the critical properties associated with the phase
transition surface $T_{c}\left( x,y,M\right) $ appear to fall into
the 3D-XY universality class\cite {book,tsparks}. Supposing that
the 2D-XY-QSI and 3D-XY universality classes hold true, this leads
naturally to universal critical amplitude combinations, involving
the transition temperature and the critical amplitudes of specific
heat and penetration depths, which also apply to the isotope
effects. They yield universal relations between the isotope
coefficients for the transition temperature, the penetration
depths and the specific heat.

In Sec.II we sketch the universal relations for anisotropic
superconductors falling at finite temperature into the 3D-XY- and
at zero temperature into the 2D-QSI universality class. Since
fluctuations are supposed to dominate, pair formation and pair
condensation do not occur simultaneously. For this reason, the
universal relations between transition temperature and the
critical amplitudes of specific heat, correlation lengths and
penetration depths, do not depend on the underlying pairing
mechanism, but reveal how the isotope effects on these quantities
are correlated. To explore the nature of the isotope effects we
concentrate on the regime where the materials flow to 2D-QSI
criticality, driven by variation of dopant concentration $x$ and
$y$, e.g. the concentration of substituents for Cu or the strength
of disorder. In this regime, the essential effect of isotope
substitution is a slight depression of the phase transition
surface $T_{c}\left( x,y\right) $ which unalterably implies a
shift of the quantum phase transition line $y_{c}\left( x\right) $
and of the underdoped limit $x_{u}$\cite{tshkprl,tsparks}. Using
this facts and the scaling theory of quantum critical phenomena it
is shown that the flow to 2D-QSI criticality is characterized by
an anomalous increase of the isotope coefficient for the
transition temperature $\beta _{T_{c}}$, zero temperature in-plane
penetration depth $\beta _{1/\lambda _{ab}^{2}(0)}$ and
condensation energy $\beta _{E}$, given by $\beta _{T_{c}}\propto
\beta _{1/\lambda _{ab}^{2}(0)}\propto \beta _{E}\propto
T_{cm}/T_{c}$, irrespective along which axis the 2D-QSI critical
point is approached. $T_{cm}$ denotes the maximum transition
temperature within a family of cuprates.

A comparison with experiment is presented in Sec.III. Although the
data on the isotope effect on transition temperature and the
critical amplitudes of the penetration depths are rather sparse,
they reveal suggestive consistency with this scenario. We hope
that the present paper will serve as a stimulus for future
systematic measurements. On the contrary, the experimental data on
the $T_{c}$ dependence of $\beta _{T_{c}}$ allows a rather
quantitative comparison with our predictions. Irrespective of the
path along which the 2D-QSI critical point is approached, our
prediction, $\beta _{T_{c}}\propto T_{cm}/T_{c}$ is well
confirmed, including the spin-triplet p-wave superconductor
Sr$_{2}$RuO$_{4}$. Here, reduced $T_{c}$'s were achieved by using
crystals containing impurities and defects\cite{mao}. Intriguingly
enough the factor of proportionality appears to adopt a nearly
unique value, indicating that $\beta _{T_{c}}\approx 0.17\ \
T_{cm}/T_{c}$ holds, independent of the material and the direction
of the flow to 2D-QSI criticality. A potential candidate for the
description of the 2D-QSI transition are 2D disordered bosons with
long-range Coulomb interactions\cite
{tshkprl,book,tsparks,fisher,cha,herbut}. Here the transition is
driven by the competition between exchange of pairs, which favors
superfluidity, and the combined effects of disorder and the
Coulomb interaction, which favor localization.

\section{Universal relations}

The universality class to which the cuprates at finite temperature
belong is not only characterized by its critical exponents but
also by various critical-point amplitude combinations which are
equally important. Though these amplitudes depend on $x,y$ and
$M$, their universal combinations do not. An important example is
the universal relation\cite{book,tsparks}
\begin{equation}
\left( k_{B}T_{c}\right) ^{3}=\left( \frac{\Phi _{0}^{2}}{16\pi
^{3}}\right) ^{3}\frac{\xi _{x0}^{tr}\xi _{y0}^{tr}\xi
_{z0}^{tr}}{\lambda _{x0}^{2}\lambda _{y0}^{2}\lambda _{z0}^{2}},
\label{eq1}
\end{equation}
where $\xi _{i0}^{tr}$ and $\lambda _{i0}$ are the critical
amplitude of the transverse correlation lengths and the
penetration depths, respectively. These length scales diverge as
\begin{equation}
\xi _{i}^{tr}=\xi _{i0}^{tr}\left| t\right| ^{-\nu }\text{, }\ \
\lambda _{i}=\lambda _{i0}\left| t\right| ^{-\nu /2}.  \label{eq2}
\end{equation}
Close to criticality the specific heat singularity is
characterized by
\begin{equation}
\frac{C}{Vk_{B}}=-\frac{T}{k_{B}}\frac{\partial
^{2}f_{s}}{\partial t^{2}}\approx \frac{A^{\pm }}{\alpha }\left|
t\right| ^{-\alpha }, \label{eq3}
\end{equation}
and the combination of the critical amplitude $A^{-}$ and the
transverse correlation volume
\begin{equation}
\left( R^{\pm }\right) ^{3}=A^{-}\xi _{x0}^{tr}\xi _{y0}^{tr}\xi
_{z0}^{tr}=-Q_{3}^{\pm }\alpha \left( 1-\alpha \right) \left(
2-\alpha \right) ,  \label{eq4}
\end{equation}
turns out to be universal as well\cite{book,tsparks}. $R^{\pm }$
are universal numbers. Combining then the universal relations
(\ref{eq1}) and (\ref{eq4}) we obtain a relation between
measurable properties which holds at any finite temperature
irrespective of the doping and substitution level:
\begin{equation}
T_{c}^{3}=\left( \frac{\Phi _{0}^{2}}{16\pi ^{3}k_{B}}\right)
^{3}\frac{\left( R^{-}\right) ^{3}}{A^{-}\lambda _{x,0}^{2}\lambda
_{y,0}^{2}\lambda _{z,0}^{2}}  \label{eq5}
\end{equation}
Considering variations in the dopant, substituent concentration
etc., denoted by $y$ it then follows that the changes in $T_{c}$,
$A^{-}$ and $1/\lambda _{i,0}^{2}$ are not independent but related
by
\begin{equation}
\frac{1}{T_{c}}\frac{dT_{c}}{dy}=-\frac{1}{3A^{-}}\frac{dA^{-}}{dy}+\frac{1}{3}\sum_{i=1}^{3}\lambda
_{i,0}^{2}\frac{d\left( 1/\lambda _{i,0}^{2}\right) }{dy}.
\label{eq6}
\end{equation}
In tetragonal cuprates, where $\lambda _{ab0}=\lambda
_{a0}=\lambda _{b0}$, it reduces to
\begin{equation}
\frac{1}{T_{c}}\frac{dT_{c}}{dy}=-\frac{1}{3A^{-}}\frac{dA^{-}}{dy}+\frac{2}{3}\lambda
_{ab,0}^{2}\frac{d\left( 1/\lambda _{ab,0}^{2}\right)
}{dy}+\frac{1}{3}\lambda _{c,0}^{2}\frac{d\left( 1/\lambda
_{c,0}^{2}\right) }{dy} \label{eq6a}
\end{equation}

In the underdoped regime, tuned by variation of the dopant,
substitution or impurity concentration the cuprates undergo a
2D-QSI transition. Here the universal relation\cite{book,tsparks}
\begin{equation}
T_{c}=\frac{\Phi _{0}^{2}\overline{R}_{2}d_{s}}{16\pi
^{3}k_{B}}\frac{1}{\lambda _{ab}^{2}\left( 0\right) },
\label{eq6b}
\end{equation}
between transition temperature and zero temperature in-plane
penetration depth holds. $d_{s}$ denotes the thickness of the
superconducting sheets and $\overline{R}_{2}$ is a universal
dimensionless constant. Considering variations in the dopant,
substituent concentration etc., denoted by $y$ it then follows
that the changes in $T_{c}$, $d_{s}$ and $1/\lambda
_{ab}^{2}\left( 0\right) $ are not independent but related by
\begin{equation}
\frac{1}{T_{c}}\frac{dT_{c}}{dy}=\frac{1}{d_{s}}\frac{dd_{s}}{dy}+\lambda
_{ab}^{2}\left( 0\right) \frac{d}{dy}\left( \frac{1}{\lambda
_{ab}^{2}\left( 0\right) }\right) .  \label{eq6c}
\end{equation}
Thus, the approach to 2D-QSI criticality is associated with a flow
to
\begin{equation}
\frac{1}{T_{c}}\frac{dT_{c}}{dy}\rightarrow
\frac{1}{d_{s}}\frac{dd_{s}}{dy}+\lambda
_{ab0}^{2}\frac{d}{dy}\left( \frac{1}{\lambda _{ab0}^{2}}\right)
\rightarrow \frac{1}{d_{s}}\frac{dd_{s}}{dy}+\lambda
_{ab}^{2}\left( 0\right) \frac{d}{dy}\left( \frac{1}{\lambda
_{ab}^{2}\left( 0\right) } \right) ,  \label{eq6d}
\end{equation}

As the isotope substitution is concerned it is useful to rewrite
Eq.(\ref {eq6a}) in the form
\begin{equation}
\beta _{_{T_{c}}}=\frac{2}{3}\beta _{1/\lambda
_{ab,0}^{2}}+\frac{1}{3}\beta _{1/\lambda
_{c,0}^{2}}-\frac{1}{3}\beta _{A^{-}},  \label{eq7}
\end{equation}
where $\beta _{B}$ is the isotope coefficient defined by
\begin{equation}
\beta _{B}=-\frac{M}{B}\frac{dB}{dM}.  \label{eq7a}
\end{equation}
$B=(T_{c},\ \lambda _{ab,0}^{2},\ \lambda _{c,0}^{2},\ A^{-})$ and
$dM=\ ^{16}M-\ ^{18}M$. It reveals that a change of the transition
temperature upon isotope substitution is unalterably linked to a
change in the specific heat and the penetration depths. Note that
a vanishing isotope effect on $T_{c}$ merely implies that the
contributions of specific heat and penetration depths cancel. In
this context it is important to recognize that the universal
relations, including Eq.(\ref{eq5}), (\ref{eq6}) and (\ref{eq7})
also apply in the presence of disorder. Indeed, since the critical
exponent $\alpha $ of the specific heat is negative in the 3D-XY
universality class, Harris criterion implies that disorder is
irrelevant at finite temperature\cite{ma}. On the other hand,
close to 2D-QSI criticality the isotope coefficients of $T_{c}$
and zero temperature in-plane penetration depth are according to
Eq.(\ref{eq6c}) related by
\begin{equation}
\beta _{T_{c}}=\beta _{d_{s}}+\beta _{1/\lambda _{ab}^{2}\left(
0\right) }. \label{eq8}
\end{equation}
In analogy to Eq.(\ref{eq6d}), the approach to 2D-QSI criticality
is then associated with the flow

\begin{equation}
\beta _{_{T_{c}}}=\frac{2}{3}\beta _{1/\lambda
_{ab,0}^{2}}+\frac{1}{3}\beta _{1/\lambda
_{c,0}^{2}}-\frac{1}{3}\beta _{A^{-}}\rightarrow \beta
_{d_{s}}+\beta _{1/\lambda _{ab}^{2}\left( 0\right) }  \label{eq9}
\end{equation}

The 2D-QSI transition is also characterized by the scaling
relations\cite {book,tsparks}
\begin{equation}
T_{c}=a_{\delta }\delta ^{z\overline{\nu }},\ \ \overline{\xi
}_{ab}=\overline{\xi }_{ab,\delta }\delta ^{-\overline{\nu }}.
\label{eq10a}
\end{equation}
$\overline{\xi }_{ab}$ is the in-plane correlation length
extrapolated to zero temperature and $\delta $ measures the
distance from the critical point with dynamic critical exponent
$z$ and correlation length exponent $\overline{\nu }$. In cuprates
there is consistent evidence for a 2D-QSI transition with $z=1$
and $\overline{\nu }\approx 1$\cite{book,tsparks}. These estimates
coincide with the theoretical prediction for a 2D disordered
bosonic system with long-range Coulomb
interactions\cite{fisher,cha,herbut}. This strongly suggests that
in cuprate superconductors the loss of phase coherence, due to the
localization of Cooper pairs, is responsible for the 2D-QSI
transition.

To explore the scaling behavior of the isotope coefficient for the
transition temperature, $\beta _{T_{c}}$, along the hole
concentration axis $x$ we can use the empirical relation
\begin{equation}
T_{c}\left( x\right) =\frac{2T_{cm}}{x_{m}^{2}}\left(
x-x_{u}\right) \left( x_{o}-x\right) ,\ \ \
x_{m}=\frac{x_{u}+x_{o}}{2},  \label{eq10c}
\end{equation}
due to Presland \emph{et al.}\cite{presland}. It describes the
phase transition line of cuprate superconductors in the
temperature-dopant concentration ($x$) plane very well
\cite{tallon}. The hole concentration $p$ can be varied
effectively by changing the oxygen content and substitution in non
copper sites. $T_{c}$ adopts its maximum value $T_{cm}$ at optimum
doping $x=x_{m}$. $x_{u}$ denotes the underdoped and $x_{o}$ the
overdoped limit. Note that this empirical relation is fully
consistent with a 2D-QSI transition with $z\overline{\nu }=1$ and
$\delta =\left( x-x_{u}\right) /x_{u}$, because $T_{c}\left(
x\right) =\left( 4T_{cm}/\left( x_{o}-x_{u}\right) \right) \left(
x-x_{u}\right) \approx T_{cm}\left( x-x_{u}\right) $. The main
effect of isotope substitution is the shift of $x_{u}$ to a
slightly larger and of $T_{cm}$ to a slightly lower value.
Concentrating on the behavior close to 2D-QSI criticality, we
obtain with Eq.(\ref{eq7a}) the expression
\begin{equation}
\beta _{T_{c}}=\frac{4x_{u}T_{cm}}{\left( x_{o}-x_{u}\right)
T_{c}}\beta _{x_{u}}\approx \frac{T_{cm}}{T_{c}}\beta _{x_{u}},\ \
\beta _{x_{u}}=\frac{M}{x_{u}}\frac{dx_{u}}{dM}.  \label{eq10d}
\end{equation}
Thus, the flow to 2D-QSI criticality along the hole concentration
axis is characterized by the depression of $T_{c}$, giving rise to
a divergent isotope coefficient $\beta _{T_{c}}$. The amplitude of
the divergence is controlled by $\beta _{p_{u}}$, the isotope
coefficient associated with the shift of the underdoped limit upon
isotope substitution. Approaching the 2D-QSI critical point along
the disorder or substitution (e.g. substitution for Cu) axis,
denoted by $y$, $T_{c}$ vanishes as\cite{book,tsparks}
\begin{equation}
T_{c}=T_{cm}\left( 1-y/y_{c}\right) ^{z\overline{\nu }}.
\label{eq10e}
\end{equation}
Upon isotope substitution $y_{c}$ shifts to a slightly smaller
value, because $T_{c}$ is lowered (see Fig.\ref{fig1}).
Concentrating then on the divergent part of $\beta _{T_{c}}$ so
that the effect on $T_{cm\text{ }}$can be neglected we obtain with
Eq.(\ref{eq7a}) and for $z\overline{\nu }=1$ the expression
\begin{equation}
\beta _{T_{c}}=\frac{T_{cm}}{T_{c}}\beta _{y_{c}},\ \ \beta
_{y_{c}}=-\frac{M}{y_{c}}\frac{dy_{c}}{dM}.  \label{eq10f}
\end{equation}
On the other hand, approaching the 2D-QSI critical point along the
residual resistivity axis, there is the scaling
relation\cite{book,tsparks}

\begin{equation}
T_{c}=T_{cm}\left( 1-\rho _{ab}/\rho _{ab}^{c}\right) ^{z\
\overline{\nu }}, \label{eq10g}
\end{equation}
where $\rho _{ab}$ and $\rho _{ab}^{c}$ denote the residual and
critical residual in-plane resistivity, respectively. In the 2D
limit the relevant measure of resistivity is the sheet resistance
$\rho ^{\square }=d_{s}\rho _{ab}$, adopting at criticality a
fixed value, $\rho ^{\square }\approx h/\left( 4e^{2}\right)
$\cite{book,tsparks,fisher,cha}. Considering $z\overline{\nu }=1$
and approaching 2D-QSI criticality along the in-plane resistivity
axis, the isotope coefficient $\beta _{T_{c}}$ adopts then with
Eq.(\ref{eq7a}) the form
\begin{equation}
\beta _{T_{c}}=\frac{T_{cm}}{T_{c}}\frac{M}{\rho
_{ab}^{c}}\frac{d\rho _{ab}^{c}}{dM}=\frac{T_{cm}}{T_{c}}\beta
_{d_{s}},\ \ \beta _{d_{s}}=-\frac{M}{d_{s}}\frac{dd_{s}}{dM}{.}
\label{eq10h}
\end{equation}
$\beta _{d_{s}}$ denotes the isotope coefficient of $d_{s}$, the
thickness of the 2D slab. Since $d_{s}$ remains finite upon
isotope substitution, this also applies to $\beta _{d_{s}}$. Thus
close to 2D-QSI criticality we obtain with Eqs.(\ref{eq8}),
(\ref{eq10d}), (\ref{eq10f}) and (\ref{eq10h}) the relation
\begin{equation}
\beta _{T_{c}}=\frac{T_{cm}}{T_{c}}\beta
_{d_{s}}=\frac{T_{cm}}{T_{c}}\beta _{y_{c}}\approx
\frac{T_{cm}}{T_{c}}\beta _{x_{u}}\approx \beta _{1/\lambda
_{ab}^{2}\left( 0\right) },  \label{eq10k}
\end{equation}
revealing that $\beta _{T_{c}}\propto T_{cm}/T_{c}$ holds with a
unique factor of proportionality, irrespective of the axis the
2D-QSI critical point is approached.

Another quantity of interest is the condensation energy per unit
volume. Close to 2D-QSI and 3D-QSN criticality its singular part
scales as the inverse of the zero temperature correlation volume,
so that
\begin{equation}
-E\left( T=0\right) \propto \delta ^{\overline{\nu }\left(
D+z\right) }\propto T_{c}^{D+z},  \label{eq10l}
\end{equation}
using Eq.(\ref{eq10a}). Thus, along the quantum critical endline
$y_{c}\left( x\right) $ ( see Fig.\ref{fig1}), the condensation
energy vanishes, as it should be. Invoking Eqs.(\ref{eq7a}) and
(\ref{eq10l}) the isotope coefficient of the condensation energy
adopts then close to the 2D-QSI critical point the form
\begin{equation}
\beta _{E}=-\frac{M}{E}\frac{dE}{dM}=\left( 2+z\right) \beta
_{T_{c}}\propto \frac{T_{cm}}{T_{c}}  \label{eq10m}
\end{equation}

\section{Comparison with experiment}

We are now prepared to confront these predictions for the isotope
effect with experiment. Although the experimental data on the
isotope effect on transition temperature and penetration depth are
rather sparse, some insight can be obtained from the estimates
listed in Table I. La$_{2-x}$Sr$_{x}$CuO$_{4}$ and
Y$_{1-x}\Pr_{x}$Ba$_{2}$Cu$_{3}$O$_{7-\delta }$ undergo a doping
tuned 2D-QSI transition at $x\approx 0.05$ and $x\approx
0.55$\cite{sun}, respectively\cite{book,tsparks}. These estimates
have been derived from a fit of the experimental data for
$1/\lambda _{ab}^{2}\left( T\right) $\cite{hoferprl,khasanov} to
the leading critical behavior

\begin{equation}
\frac{1}{\lambda _{ab}^{2}}=\frac{1}{\lambda _{ab,0}^{2}}t^{2\nu
},\ \ t=1-T/T_{c},\ \ \nu =2/3.  \label{eq11}
\end{equation}

\begin{center}
\begin{tabular}{|l|l|l|l|l|}
\hline & $x$ & $T_{c}\ \ $($^{16}$O) (K) & $\Delta T_{c}/T_{c}$ &
$\lambda _{ab0}^{2}\ \Delta \left( 1/\lambda _{ab0}^{2}\right) $
\\ \hline La$_{2-x}$Sr$_{x}$CuO$_{4}$ & 0.15 & 37.2 & -0.008 &
-0.033 \\ \hline & 0.086 & 22.19 & -0.051 & -0.062 \\ \hline &
0.08 & 19.1 & -0.073 & -0.086 \\ \hline
Y$_{1-x}\Pr_{x}$Ba$_{2}$Cu$_{3}$O$_{7-\delta }$ & 0 & 91.4 &
-0.0044 & -0.025
\\ \hline
& 0.3 & 57.2 & -0.049 & -0.072 \\ \hline & 0.4 & 42.6 & -0.059 &
-0.079 \\ \hline
\end{tabular}
\end{center}

\bigskip

Table I: Estimates for $\Delta T_{c}/T_{c}$ and $\lambda
_{ab0}^{2}\ \Delta \left( 1/\lambda _{ab0}^{2}\right) $ derived
from the magnetic torque measurements of Hofer \emph{et
al.}\cite{hoferprl} on La$_{2-x}$Sr$_{x}$CuO$_{4}$ and the $\mu
$SR measurements of Khasanov \emph{et al.}\cite{khasanov} on
Y$_{1-x}\Pr_{x}$Ba$_{2}$Cu$_{3}$O$_{7-\delta }$.

\bigskip

\begin{figure}[tbp]
\centering
\includegraphics[totalheight=6cm]{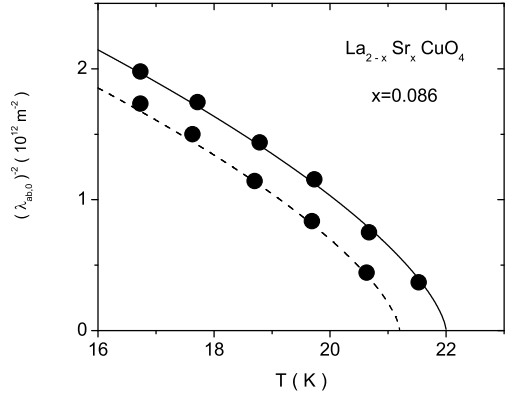}
\caption{Isotope effect on transition temperature and in-plane
penetration depth in La$_{2-x}$Sr$_{x}$CuO$_{4}$ at x=0.086. The
upper ($^{16}$O) and lower curve ($^{18}$O) are fits to
Eq.(\ref{eq11}) yielding the parameters listed in Table I.
Experimental data taken from Hofer \emph{et
al.}\protect\cite{hoferprl}.} \label{fig2}
\end{figure}

\begin{figure}[tbp]
\centering
\includegraphics[totalheight=6cm]{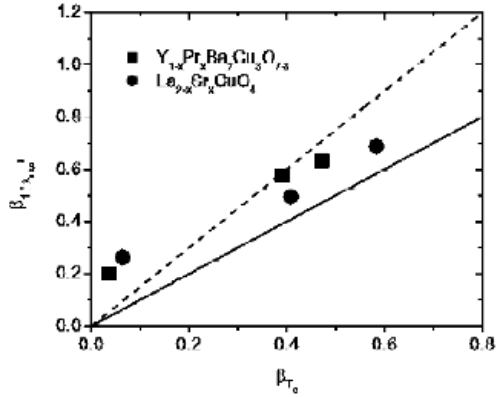}
\caption{$\beta _{1/\lambda _{ab,0}^{2}}$ versus $\beta _{T_{c}}$
for Y$_{1-x}\Pr_{x}$Cu$_{3}$O$_{7-\delta }$ ($\blacksquare :x=0,\
0.3$ and $0.4$) and La$_{2-x}$Sr$_{x}$CuO$_{4}$ ($\bullet :\ \
x=0.08,$ $0.086$ and $0.15$). The straight line is $\beta
_{1/\lambda _{ab,0}^{2}}=\beta _{T_{c}}$ and the dashed one is .
$\beta _{1/\lambda _{ab,0}^{2}}=3\beta _{T_{c}}/2$. Data points
taken from Table I.} \label{fig3}
\end{figure}

An example is shown in Fig.\ref{fig2}. Since there is mounting
evidence for 3D-XY criticality and the available data are rather
sparse close to $T_{c}$, we kept $\ \nu $ fixed. For this reason
and in view of the excellent fits to both, the
La$_{2-x}$Sr$_{x}$CuO$_{4}$ and
Y$_{1-x}\Pr_{x}$Ba$_{2}$Cu$_{3}$O$_{7-\delta }$ data, the
estimates quoted in Table I should be reliable and with that
reveal the flow to 2D-QSI criticality. To uncover this flow we
displayed in Fig.\ref{fig3} the plot $\beta _{1/\lambda
_{ab0}^{2}}\ $versus $\beta _{T_{c}}$. Although the set of data
points is sparse, key trends emerge. As $T_{c}$ decreases the data
points flow to the straight line, signaling the approach to 2D-QSI
criticality, where $\beta _{1/\lambda _{ab0}^{2}}\rightarrow \beta
_{1/\lambda _{ab}^{2}\left( 0\right) }\rightarrow \beta
_{T_{c}}\rightarrow \infty $ (see Eqs.(\ref{eq9}) and
(\ref{eq10k})). In this flow both cuprates appear to cross the
dashed line, corresponding to $\beta _{1/\lambda
_{ab0}^{2}}=3\beta _{T_{c}}/2$. Here the contributions of specific
heat and out of plane penetration depth cancel (see
Eq.(\ref{eq7})). This clearly reveals that except for this special
point the isotope effect on transition temperature and the
critical amplitude of the in-plane penetration depth are not
simply related. Indeed, the contributions of the specific heat and
out of plane penetration depth cancel at this special point only.
There is also the experimental fact that close to the maximum
$T_{c}$ the isotope effect on $1/\lambda _{ab0}^{2}$ is
substantial, while it is very small in $T_{c}$. A glance to the
universal relation (\ref{eq7}) shows that even for $\beta
_{T_{c}}=0$ a substantial isotope effect on the in-plane
penetration depth can be expected, whenever the contributions of
the specific heat and out of plane penetration depth do not
cancel. Indeed, the absence of an isotope effect in one of the
quantities entering Eq.(\ref{eq7}) is not bound to the absence of
an effect in all the others.

\begin{figure}[tbp]
\centering
\includegraphics[totalheight=7cm]{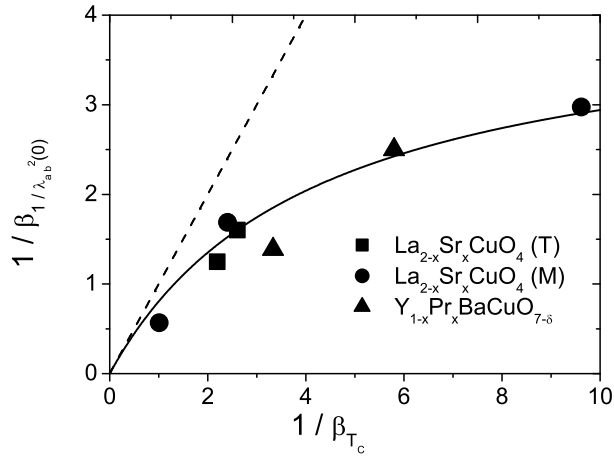}
\caption{Oxygen isotope effect in La$_{2-x}$Sr$_{x}$CuO$_{4}$ and
Y$_{1-x}\Pr_{x}$Ba$_{2}$Cu$_{3}$O$_{7}$ in terms of $\ 1/\beta
_{1/\lambda _{ab}^{2}\left( 0\right) }$ versus $1/\beta _{T_{c}}$.
The dashed line marks the critical behavior close to 2D-QSI
criticality, that is Eq.(\ref{eq8}) with $\beta _{d_{s}}=0$, while
the solid curve is Eq.(\ref{eq8}) with $\beta _{d_{s}}=-0.24$,
indicating the crossover due to the isotope effect on $d_{s} $.}
\label{fig4}
\end{figure}

Additional evidence for the flow to 2D-QSI criticality stems from
the plot $1/\beta _{1/{\lambda _{ab}^{2}(0)}}$\ versus $\ 1/\beta
_{T_{c}}$ shown in Fig.\ref{fig4} for the oxygen isotope effect in
La$_{2-x} $Sr$_{x}$Cu$_{1-x}$O$_{4}$ \cite{hoferprl,zhao} and
Y$_{1-x}\Pr_{x}$Ba$_{2}$Cu$_{3}$O$_{7}$ \cite{khasanov}. It
clearly reveals the crossover to the asymptotic 2D-QSI behavior,
indicated by the dashed straight line, where $\beta
_{T_{c}}\rightarrow \beta _{1/{\lambda _{ab}^{2}(0)}}$, because
$\beta _{d_{s}}$ remains finite, while $\beta _{T_{c}}=T_{cm}\beta
_{d_{s}}/T_{c}$ diverges (Eq.(\ref{eq10k}). For comparison we
included the solid curve, which is Eq.(\ref{eq8}), providing for
$\beta _{d_{s}}$,  the oxygen isotope coefficient of $d_{s}$, the
estimate
\begin{equation}
\beta _{d_{s}}=-\frac{M}{d_{s}}\frac{dd_{s}}{dM}=-0.24,
\label{eq11b}
\end{equation}
This yields with Eq.(\ref{eq10k})
\begin{equation}
\beta _{T_{c}}\approx 0.24\frac{T_{cm}}{T_{c}},  \label{eq11c}
\end{equation}
irrespective of the path the 2D-QSI transition is approached.

\begin{figure}[tbp]
\centering
\includegraphics[totalheight=7cm]{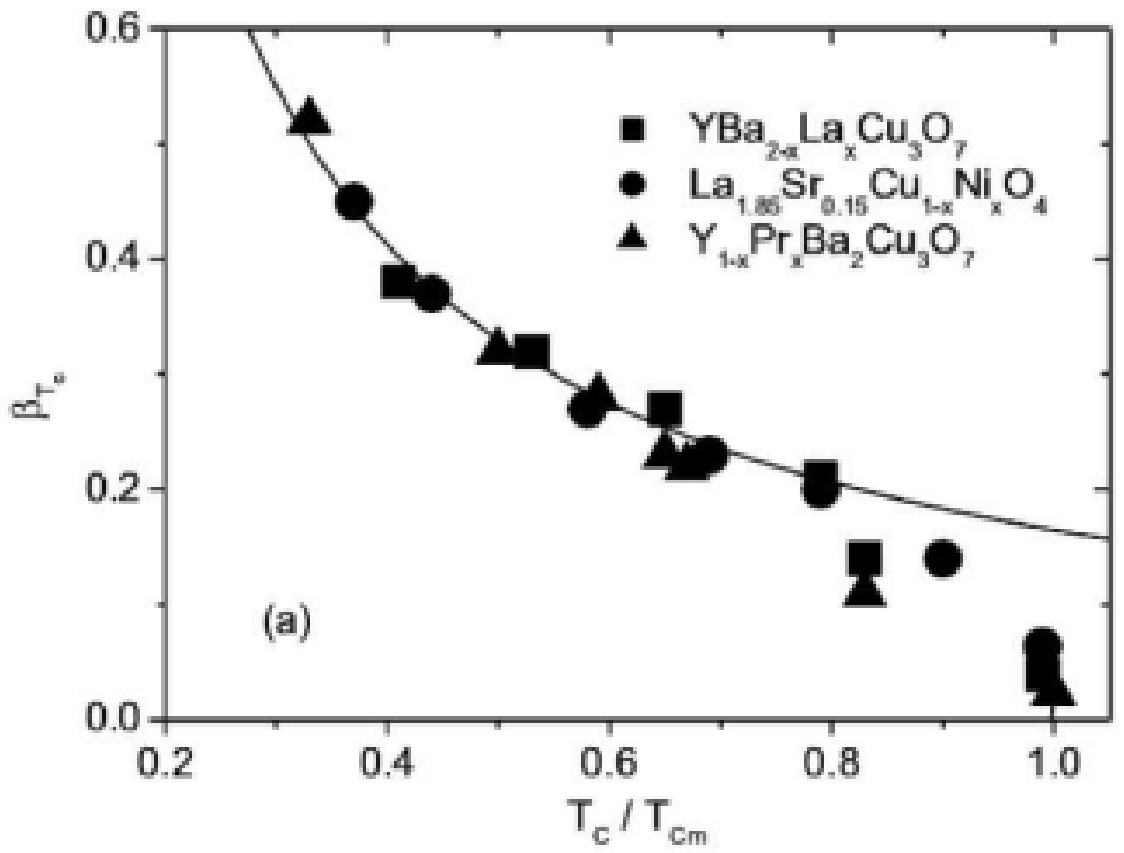}
\includegraphics[totalheight=7cm]{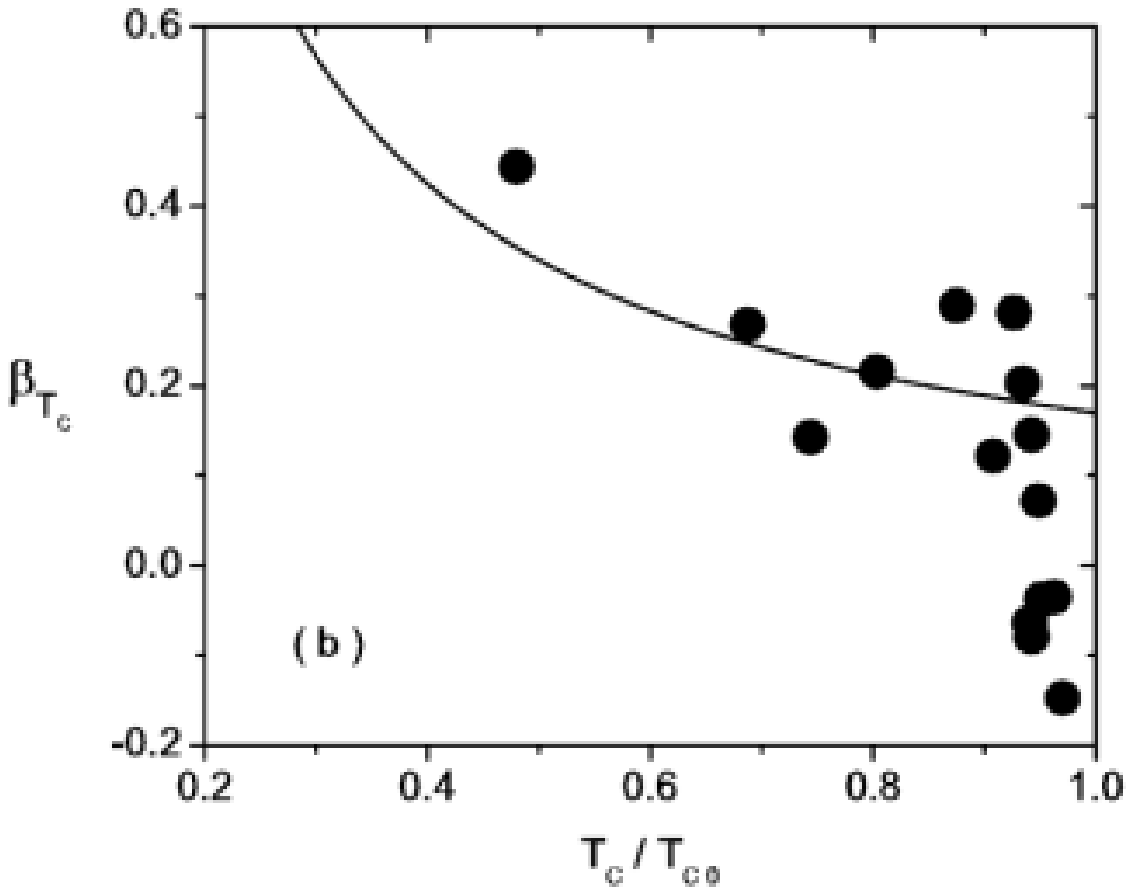}
\caption{a. $\beta _{T_{c}}$ versus $T_{c}/T_{cm}$
YBa$_{2-y}$La$_{y}$Cu$_{3}$O$_{7}$ \protect\cite{bornemann},
La$_{1.95}$Sr$_{0.15}$Cu$_{1-y}$Ni$_{y}$O$_{4}$\protect\cite{babushkina}
and Y$_{1-x}\Pr_{x}$Ba$_{2}$Cu$_{3}$O$_{7}$\protect\cite{franck}.
The solid curve is Eq.(\ref{eq12}); b. $\beta _{T_{c}}$ versus
$T_{c}/T_{c0}$ for Sr$_{2}$RuO$_{4}$. Experimental data taken from
Mao \emph{et al.}\protect\cite{mao}. The solid curve is
Eq.(\ref{eq12}).} \label{fig5}
\end{figure}

To clarify the uniqueness of this estimate for the factor of
proportionality further, we displayed in Fig.\ref{fig5} the
experimental data for
YBa$_{2-y}$La$_{y}$Cu$_{3}$O$_{7}$\cite{bornemann},
La$_{1.95}$Sr$_{0.15}$Cu$_{1-y}$Ni$_{y}$O$_{4}$\cite{babushkina},
Y$_{1-x}\Pr_{x}$Ba$_{2}$Cu$_{3}$O$_{7}$\cite{franck} and
Sr$_{2}$RuO$_{4}$\cite{mao} in terms of $\beta _{T_{c}}$ versus
$T_{c}/T_{cm}$. Note that Sr$_{2}$RuO$_{4}$ is a spin-triplet
p-wave superconductor with an intrinsic $T_{c0\text{ }}=1.5$K. It
is the only layered perovskite superconductor without Cu so far.
Reduced $T_{c}$'s were achieved by using crystals containing
impurities and defects. Moreover, in conjunction with the scaling
relations (\ref{eq10a}) and (\ref{eq10g}) the experimental data on
the $T_{c}$ dependence of the in-plane correlation length, as well
as the dependence of $T_{c}$ on residual in plane resistivity,
point to a 2D-QSI transition with $z=1$ and $\overline{\nu
}\approx 1$\cite{mao2}. Thus, in agreement with the scaling
predictions (\ref {eq10k}) and (\ref{eq11c}) for $\beta _{T_{c}}$
versus $T_{cm}/T_{c}$, $\beta _{T_{c}}$ increases with reduced
$T_{c}$ and, as indicated by the solid curves in Fig.\ref{fig5},
tends to diverge as
\begin{equation}
\beta _{T_{c}}\approx 0.17\frac{T_{cm}}{T_{c}}.  \label{eq12}
\end{equation}
Note that the copper isotope effect in
YBa$_{2}$Cu$_{3}$O$_{7-\delta }$\cite {zhao2} leads to analogous
behavior. Thus, there is the remarkable experimental fact that $\
\beta _{y_{c}}$, $\beta _{p_{u}}$ and $\beta _{d_{s}}$ describing
the shift of the critical point along the disorder, Cu
substitution, dopant and sheet resistance axis , respectively (see
Eqs.(\ref {eq10k}) and Eq.(\ref{eq11c})) adopt the nearly unique
value $\beta _{x_{u}}\approx \ \beta _{y_{c}}\approx \beta
_{d_{s}}\approx 0.17-0.24$, including the spin-triplet p-wave
superconductor Sr$_{2}$RuO$_{4}$. It stems from a mechanism which
shifts $x_{u}$, $y_{c}$ and $d_{s}$ upon isotope substitution,
subject to the intriguing constraint that the associated
coefficient $\beta _{x_{u}}$, $\beta _{y_{c}}$ and $\beta
_{d_{s}}$ adopt nearly the same value. As a consequence, in these
materials the isotope effects and the pairing mechanism are
unrelated. In this context it is important to note that the
simultaneous occurrence of pairing and pair condensation requires
the absence of thermal and quantum fluctuations. On the contrary,
we have seen that cuprates flow, tuned by the combined effect of
quantum fluctuations and disorder, to 2D-QSI criticality and that
the isotope effects on transition temperature, specific heat and
penetration depths are not independent but related by the
universal relation (\ref{eq7}) which requires thermal fluctuations
to dominate. Given this behavior it follows that along the flow to
2D-XY-QSI criticality fluctuations are relevant so that pairing
and pair condensation do not occur simultaneously. The generically
small value of $\beta _{T_{c}}$ around the maximum transition
temperature is then due to the cancellation of the specific heat
and penetration depths contributions (Eq. (\ref{eq7})), while the
rise of $\beta _{T_{c}}$ with reduced $T_{c}$ (Fig.\ref{fig4})
reflects the flow to 2D-XY-QSI criticality, where the scaling
predictions (\ref{eq10k}), (\ref {eq11c}) and (\ref{eq12}) apply.

We have seen that close to 2D-QSI criticality the isotope effects
stem from the shift of the phase diagram and $d_{s}$ upon isotope
substitution. On the other hand, the 2D-QSI transition is driven
by the exchange of pairs which favors superconductivity and the
combined effects of disorder and Coulomb repulsion of the pairs,
which favor localization. Since isotope substitution will hardly
affect disorder and Coulomb repulsion, the mechanism left is the
electron-phonon interaction, which renormalizes the mass of the
fermions and with that mass and exchange of the pairs. Indeed,
inelastic neutron-scattering experiments on cuprate
superconductors give clear evidence of a strong electron-phonon
coupling \cite{pintschovius}. In particular, characteristic phonon
anomalies generated by a large electron-lattice interaction have
been detected in La$_{2}$CuO$_{4}$
\cite{pintschovius,pintschovius,queeney,reichardt}. Here a strong
softening of the high-frequency copper-in-plane-oxygen
bond-stretching modes is found as holes are doped in the
insulating parent compound, i.e., we have an unmistakable and
important example for a direct correlation of hole doping with the
phonon anomalies. Doping leads to a strong decrease of the planar
oxygen breathing mode frequency and even more strongly for a CuO
bond-stretching vibration. Similar to the observations in doped
La$_{2}$CuO$_{4}$ pronounced phonon anomalies for the
Cu-in-plane-oxygen bond stretching vibrations have also been
observed by neutron scattering experiments in
YBa$_{2}$Cu$_{3}$O$_{7}$\cite{pintschovius2,reichardt}. These
experimental results strongly suggest that the anomalous phonon
softening upon hole doping and the associated strong
electron-phonon coupling is a generic feature of cuprates.
Moreover the measured frequency shift of the transverse optic
phonons due to the substitution of $^{16}$O by $^{18}$O yielded in
underdoped YBa$_{2}$Cu$_{3}$O$_{7}$ ($T_{c}=68\ $K) an isotope
coefficient of the expected magnitude for copper-oxygen stretching
modes, with $\beta _{\omega }=0.5\pm 0.1$\cite{wang}

\section{Summary and conclusions}

We have seen that the experimental data on the isotope effects on
transition temperature and in-plane penetration depth in cuprate
superconductors reveal remarkable consistency with the universal
relations for a system undergoing a 3D-XY phase transition at
finite temperature and a quantum superconductor to insulator
transitions in two dimensions (2D-QSI). Although the relevant
experimental data on the effect on transition temperature and
in-plane penetration depth are rather sparse and the effect on the
out of plane penetration depth and specific heat remains to be
investigated, there is mounting evidence that cuprates fall at
finite temperature into the 3D-XY- and at zero temperature, close
to the insulator-superconductor boundary into the 2D-XY-QSI
universality classes. By definition, the universal relations hold
irrespective of the dopant or substituent concentration. They
apply whenever critical fluctuations dominate. In this case
pairing and pair condensation do not occur simultaneously. This
differs from conventional superconductors where fluctuations do
not play an essential role and mean-field treatments appear to
work. Here pairing and pair condensation occurs simultaneously and
due to the absence of fluctuations the universal relations do not
apply. Given then the analysis presented here, together with
additional evidence for the importance of critical fluctuations in
the cuprates, it becomes clear, that the isotope effects in these
materials do not provide direct information on the pairing
mechanism. These effects are subject to the universal relations
and the phase diagram. In the underdoped cuprates considered here,
they provide useful information on the flow to 2D-QSI criticality
and the interplay between the isotope effects on transition
temperature, specific heat and penetration depths. Concerning the
mechanism of the isotope effects, it leads to a shift of the phase
diagram. $T_{c}$ is lowered and this implies unalterably a shift
of the critical dopant, substituent, impurity and defect
concentration, where the 2D-QSI transition occurs. This transition
is driven by the exchange of pairs, which favors superconductivity
and the combined effects of disorder and Coulomb repulsion of the
pairs, which favor localization. Since isotope substitution will
hardly affect disorder and Coulomb repulsion, it appears most
likely that electron - phonon interaction, which renormalizes the
mass of the fermions and with that the mass and the exchange of
the pairs, is the relevant mechanism, giving rise to the shift of
the phase diagram.

Finally we hope that this novel point of view about the isotope
effects in fluctuation dominated superconductors will stimulate
further experimental work to obtain new data to confirm or refute
our predictions.

The author is grateful to D. Di Castro, R. Khasanov, H. Keller ,
K. A. M\"{u}ller and J. Roos for stimulating discussions.

\end{document}